\documentclass[11pt]{article}
\usepackage{graphicx}
\usepackage{epsf}  
\usepackage{epsfig}  

\newcommand{\BABARPubYear}    {01}

\newcommand{\BABARConfNumber} {06}
\newcommand{\SLACPubNumber} {8914}

\newcommand{\psfiletwoBB}[5]{ 
  \begin{minipage}{\linewidth}
    \parbox[b]{.49\linewidth}{%
      \begin{center}
        \setlength{\epsfxsize}{#5\linewidth}\leavevmode\epsfbox[#1]{#2}
      \end{center}
    }
    \hfill
    \parbox[b]{.49\linewidth}{%
      \begin{center}
        \setlength{\epsfxsize}{#5\linewidth}\leavevmode\epsfbox[#3]{#4}
      \end{center}
    }
  \end{minipage}
}
\newcommand{\psfile}[3][]{ 
  \begin{center}
    \setlength{\epsfxsize}{#3\linewidth}\leavevmode
    \def\noOpt{}\def\testit{#1}\ifx\testit\noOpt%
      \epsfbox{#2}%
    \else%
      \epsfbox[#1]{#2}%
    \fi
  \end{center}
}

\input pubboard/babarsym
\def\dbline{\noalign{\vskip 0.15truecm\hrule}\noalign{\vskip 2pt}\noalign{\hrule\vskip 0.15truecm}}
\def\sgline{\noalign{\vskip 0.20truecm\hrule\vskip 0.20truecm}}

\newcommand{\calB}{\mbox{${\cal B}$}}
\newcommand{\DE}{\ensuremath{\Delta E}}
\newcommand{\xf}{\mbox{${\cal F}$}}

\newcommand{\etaKst}{\mbox{$B\ra\eta K^*$}}
\newcommand{\etaKstp}{\mbox{$B^+\ra\eta K^{*+}$}}
\newcommand{\fetaKstp}{\mbox{$\eta K^{*+}$}}

\newcommand{\etaKstz}{\mbox{$B^0\ra\eta K^{*0}$}}
\newcommand{\fetaKstz}{\mbox{$\eta K^{*0}$}}
\newcommand{\BetaKstp}{\mbox{$\calB(B^+\ra\eta K^{*+})$}}
\newcommand{\BetaKstz}{\mbox{$\calB(B^0\ra\eta K^{*0})$}}

\newcommand{\etagg}{\mbox{$\eta\ra\gaga$}}

\newcommand{\kstzd}{\mbox{$K^{*0}\ra K^+\pi^-$}}

\newcommand{\kstpkz}{\mbox{$K^{*+}\ra \KS\pi^+$}}
\newcommand{\kshortpipi}{\mbox{$\KS\ra \pi^+\pi^-$}}

\long\def\inst#1{\par\nobreak\kern 4pt\nobreak
    {\it #1}\par\vskip 10pt plus 3pt minus 3pt}

\begin{document}
{\pagestyle{empty}

\begin{flushright}
\babar-CONF-\BABARPubYear/\BABARConfNumber \\
SLAC-PUB-\SLACPubNumber \\
July, 2001 \\
\end{flushright}

\par\vskip 5cm

\begin{center}
\Large \bf Measurement of the exclusive branching fractions \etaKstz\ and \etaKstp.
\end{center}
\bigskip

\begin{center}
\large The \babar\ Collaboration\\
\mbox{ }\\
\today
\end{center}
\bigskip \bigskip

\begin{center}
\large \bf Abstract
\end{center}
We present the results of searches for $B$ decays to the two charmless two-body final states
\etaKstz and \etaKstp, based on 20.7 \invfb\ of data collected in 1999 and 2000 with the \babar\ detector
at \pep2.
We find the branching fractions $\BetaKstz = (19.8^{+6.5}_{-5.6}\pm1.7)\times 10^{-6}$
and $\BetaKstp\ = (22.1^{+11.1}_{-9.2}\pm3.3)\times 10^{-6}$, where the first error
quoted is the statistical and the second systematic.

\vfill
\begin{center}
Submitted to the International Europhysics 
Conference on High Energy Physics, \\
7/12---7/18/2001, Budapest, Hungary
\end{center}

\vspace{1.0cm}
\begin{center}
{\em Stanford Linear Accelerator Center, Stanford University, 
Stanford, CA 94309} \\ \vspace{0.1cm}\hrule\vspace{0.1cm}
Work supported in part by Department of Energy contract DE-AC03-76SF00515.
\end{center}

}
\newpage

\begin{center}
\small

The \babar\ Collaboration,
\bigskip

B.~Aubert,
D.~Boutigny,
J.-M.~Gaillard,
A.~Hicheur,
Y.~Karyotakis,
J.~P.~Lees,
P.~Robbe,
V.~Tisserand
\inst{Laboratoire de Physique des Particules, F-74941 Annecy-le-Vieux, France }
A.~Palano
\inst{Universit\`a di Bari, Dipartimento di Fisica and INFN, I-70126 Bari, Italy }
G.~P.~Chen,
J.~C.~Chen,
N.~D.~Qi,
G.~Rong,
P.~Wang,
Y.~S.~Zhu
\inst{Institute of High Energy Physics, Beijing 100039, China }
% ...

\end{center}\newpage

\section{Introduction}
\label{sec:Introduction}
We report results for searches for $B$ decays to the charmless two-body final states
\etaKstz\ and \etaKstp\ \cite{bib:conjugate}.
These processes are manifestations of penguin or suppressed tree amplitudes proportional
to small couplings in hadronic flavor mixing (CKM matrix \cite{CKM}). As more of these rare
decay modes are measured, their phenomenological description will improve, and with it the sensitivity
to any contributions through virtual particle loops or interference terms of heretofore undetected physics.

\section{The \babar\ detector and dataset}
\label{sec:babar}
The data were collected with the \babar\ detector \cite{babar}
at the \pep2\ storage ring~\cite{pep} located at the Stanford Linear Accelerator Center. 
The results presented in this paper are based on data taken
in the 1999--2000 run. An integrated
luminosity of 20.7~fb$^{-1}$ was recorded at the $\Upsilon (4S)$ resonance
corresponding to 22.7 million $B\overline{B}$ pairs 
(``on-resonance''). In addition 2.6~fb$^{-1}$ was recorded about 40~\mev\ below
this energy (``off-resonance'') to study non-\bbbar\ continuum.

The asymmetric beam configuration in the laboratory frame
provides a boost to the $\Upsilon(4S)$
increasing the momentum range of the $B$-meson decay products
up to 4.3~\gevc.
Charged particles are detected and their momenta are measured
by a combination of a silicon vertex tracker (SVT) consisting
of five double-sided layers and a 40-layer drift chamber
(DCH), both operating in a 1.5~T solenoidal magnetic field.
Photons are detected by a CsI electromagnetic calorimeter (EMC), which
provides excellent angular and energy resolution with high efficiency for
energies above 20~\mev.

Charged particle identification (PID) is provided by the specific ionization 
loss (\dedx) in the tracking devices  and
by a unique, internally reflecting ring imaging
Cherenkov detector (DIRC) covering the central region.
A Cherenkov angle $K$--$\pi$ separation of better than 4$\sigma$ is
achieved for tracks below 3~\gevc\ momentum, decreasing to
2.5$\sigma$ at the highest momenta in our final states \cite{KpiPRL}.

\section{Analysis method}
\label{sec:Analysis}

We reconstruct a $B$ meson candidate by combining an $\eta$
candidate with a \Kstar\ candidate. The daughter resonance decays are \etagg,
\kstzd, \kstpkz\ and \kshortpipi.
These modes are kinematically distinct from the dominant $B$ decays to heavier
charmed daughters.  Backgrounds come primarily from combinatorics among
continuum events in which a light quark pair was produced instead of an
\FourS.

Monte Carlo (MC) simulations \cite{geant}\ of the target decay modes and
of continuum background were used to establish the event selection criteria.
They were designed to achieve high
efficiency and retain sidebands sufficient to characterize the
background for subsequent fitting.
Photons must satisfy $E_\gamma>$ 50 \mev\ for $\eta$ candidates.
We select $\eta$ and \Kstar\ candidates with the requirements
$490 < m_{\gamma\gamma} < 600$ \mevcc, and $800< m_{K\pi} <990$
\mevcc.  For \KS\ candidates we require
$400 < m_{\pi\pi} < 600$ \mevcc.

The pion (kaon) daughters of the \Kstar\ candidates must have DIRC, \dedx, and EMC
responses consistent with pions (kaons). 
For the \KS, the three-dimensional flight distance from the event primary
vertex must exceed 2 mm, the two-dimensional angle between the flight and momentum vectors
must be less than 40 mrad and the lifetime significance ($\tau/\sigma_\tau$) should be larger than 3.

A $B$ meson candidate is characterized by two kinematic observables.
In the CMS system, due to the two-body nature of the $B$ meson production at the \FourS,
the $B$ meson candidate's energy $E^*_B$ must be equal to $\sqrt{s}/2$, where $\sqrt{s}$ is the center of
mass energy.
This is taken into account by requiring that $|\DE| = |E^*_B - \sqrt{s}/2|$ be less than 0.2 \gev,
and the beam energy constrained mass $\mec = \sqrt{\hat{E}^{*2}_B - \hat{p}^{*2}_B}$, where $\hat{E}^*_B$ and
$\hat{p}^*_B$ are values obtained from a kinematic fit with the constraint $E^*_B = \sqrt{s}/2$.

To discriminate against tau-pair and two-photon background we require the
event to contain at least three (four) charged tracks for neutral (charged) $B$ meson candidates.
To reject continuum background we make use of the angle $\theta_T$ between the thrust axis
of the $B$ candidate and the rest of the tracks and neutral clusters in
the event, calculated in the center-of-mass frame.  The distribution of
$\cos{\theta_T}$ is sharply peaked near
$\pm1$ for combinations drawn from jetlike $q\bar q$ pairs, and nearly
uniform for the isotropic $B$ meson decays.
We require $|\cos{\theta_T}|\le0.9$.

Event yields are obtained by an unbinned extended maximum likelihood (ML) fit
analysis, while requirement based analyses are used to validate the results.
The input observables are \DE, \mec, the invariant masses $m_{\gamma\gamma}$ and $m_{K\pi}$
of the two resonant daughter candidates and a Fisher discriminant \xf.
The \KS\ spectrum is not fitted because candidates in the background are dominantly real \KS.
The Fisher discriminant \cite{CLEO-fisher}\ combines two production angles and a nine bin
representation of the energy flow about the $B$ decay axis. 
For the $\eta$ mode the helicity angle $\theta^{\rm hel}_\eta$ is the angle in the
$\eta$ rest frame between the direction of one of the photons and the $\eta$ flight direction.
We require $\cos\theta^{\rm hel}_{\eta} \le 0.92$ to discriminate against $\Kstar\gamma$ background.
A second $B$ candidate satisfying the preliminary requirements occurs in about
$11\%$ of the events.  In this case the ``best'' combination is selected
according to a $\chi^2$ computed from $m_\eta$ and $m_{\Kstar}$.

The requirement based analyses use the same variables as the ML fit with tighter
selection criteria for the signal.  A large sideband in the \mes, \DE\
plane gives an estimate of the continuum background which, with
appropriate scaling, is subtracted from the raw signal yield.

We use MC to estimate backgrounds from other $B$ decays,
including modes with and without charmed daughters.  We find these contributions
to be negligible.

The likelihood function for $N$ observed events is
\begin{displaymath}
 {\cal L} = \frac{e^{-(\sum n_j)}}{N!} \prod_{i=1}^N {\cal L}_i\,,
\end{displaymath}
where the contribution of event $i$ is
\begin{displaymath}
{\cal L}_i = \sum_{j=1}^{m}n_j {\cal P}_j(\vec{x}_i).
\label{eq:evtL}
\end{displaymath}
Here $n_j$ is the population size for species $j$ (e.g., signal,
background) and ${\cal P}_j(\vec{x}_i)$ the corresponding probability
distribution function (PDF),
evaluated with the observables $\vec{x}_i$ of the $i$th event.

For the fits ${\cal L}_i$ becomes
(with the event index $i$ suppressed on both sides of the equation) $$
{\cal L} = n_S {\cal P}_{S} +  n_C {\cal P}_{C} \,,
$$
where $n_S$ is the number of signal events and $n_C$ is the number of continuum background
events. These quantities are the free parameters of the ML
fit. The probabilities for the components are ${\cal P}_{S}$ for signal and ${\cal P}_{C}$
for background.
Since we measure the correlations among the observables in the data to
be small, we take each ${\cal P}_j$ to be a product of the PDFs for the
separate observables. 

We determine the PDFs for the likelihood fit from
simulation for the signal component, and off-resonance and on-resonance sideband
data for the continuum background.  Peaking distributions (signal
masses, \DE, \xf) are parameterized as ``crystal ball shape'' \cite{cbshape}, double Gaussian or bifurcated Gaussian functions.
Slowly varying distributions (combinatoric background under mass
or energy peaks) have polynomial shapes. 
The combinatoric background in \mec\ is described by a phase
space motivated empirical function \cite{argus}, the Argus shape. 
Control samples of $B$ decays to charmed final states of similar topology
are used to verify the simulated resolutions in \DE\
and \mec.  

\section{Results}
\label{sec:Physics}

\begin{figure}[htbp]
 \psfiletwoBB{75 155 540 610}{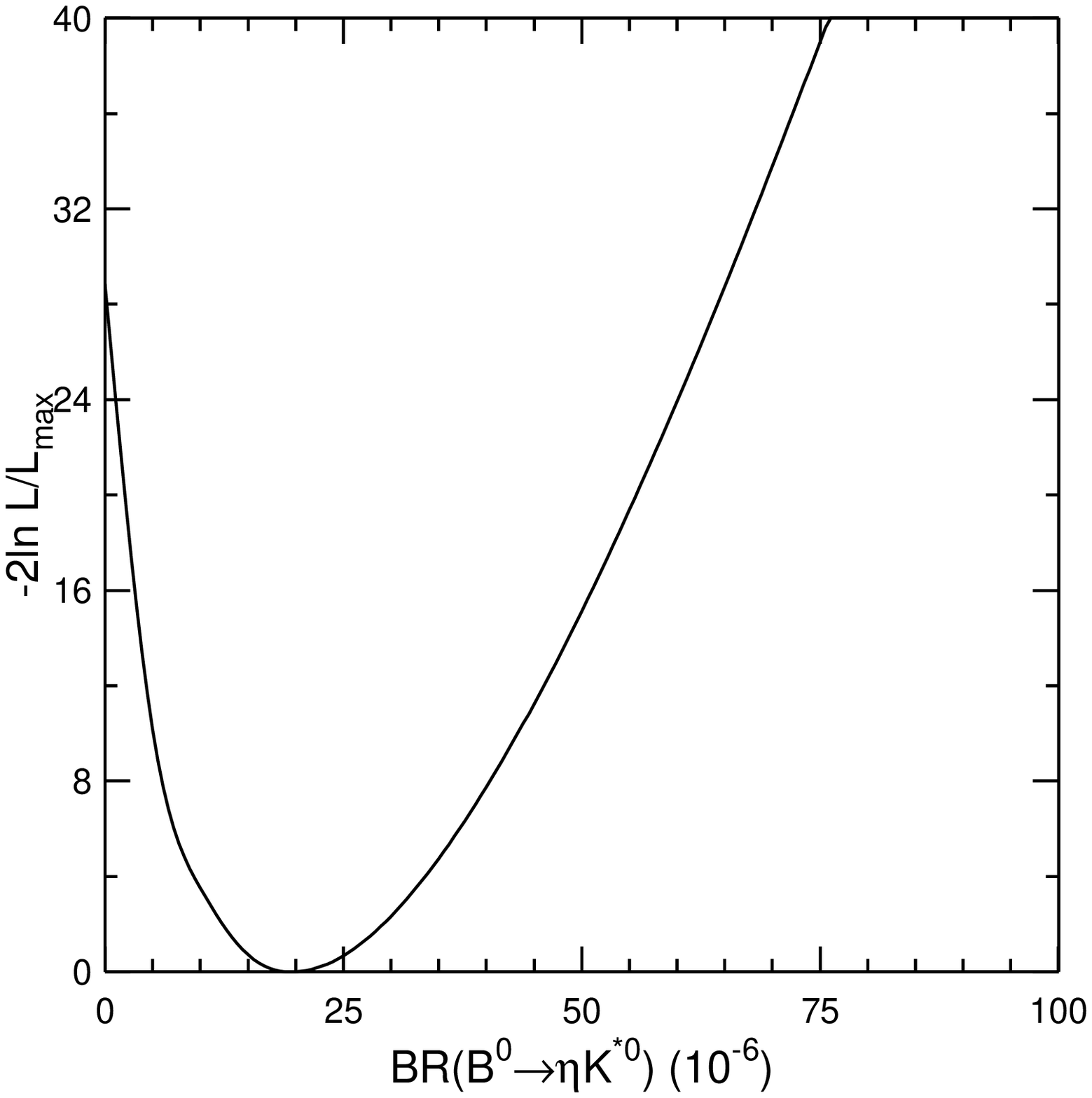}{75 155 540 610}{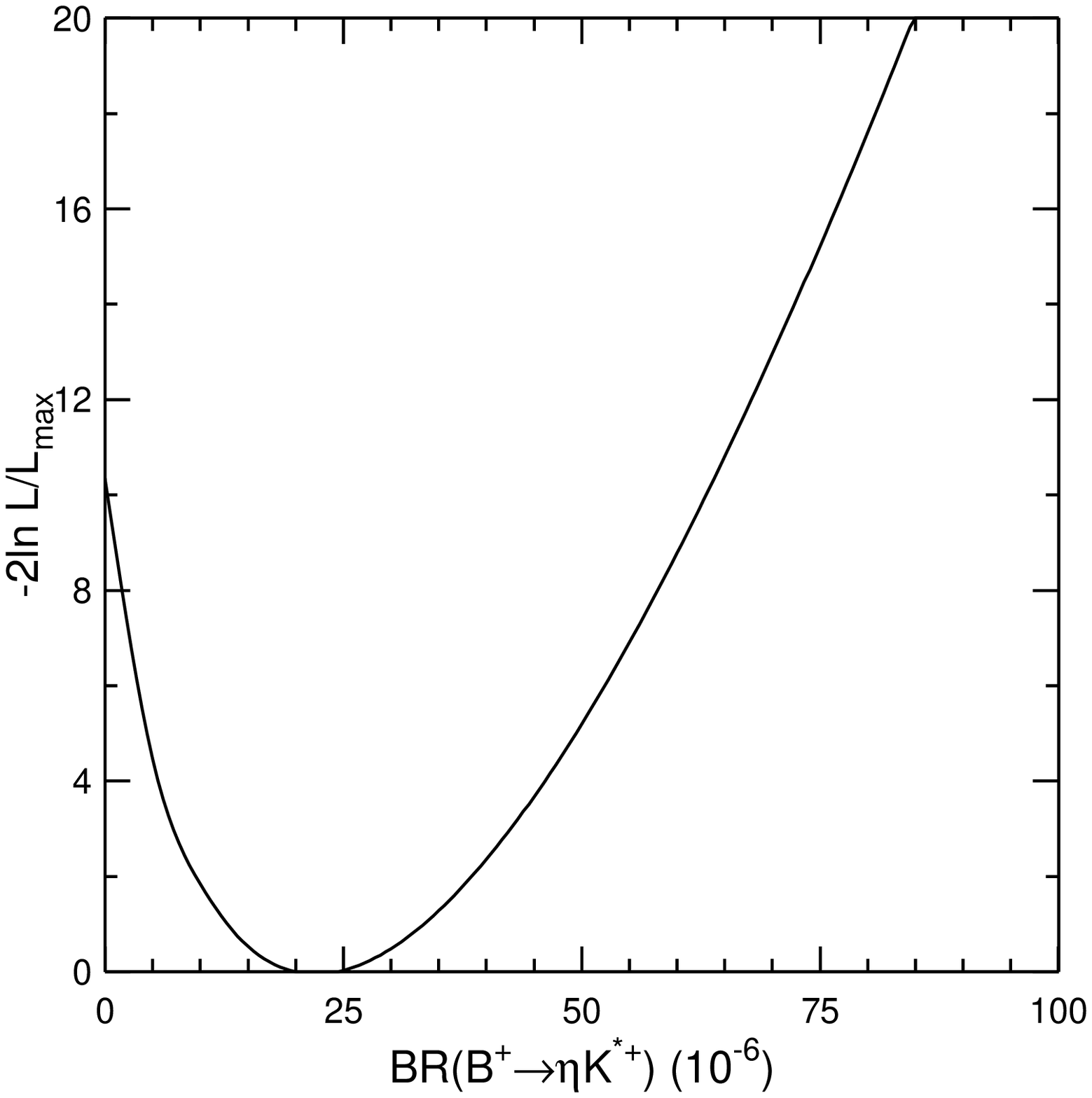}{0.9}
\vspace{-0.5cm}
 \caption{\label{fig:Cchisq}
likelihood functions for  \etaKstz\ (left) and \etaKstp\ (right) branching fractions.}
\end{figure}

We compute the branching fractions from the fitted signal event yields,
reconstruction efficiency, daughter branching fractions, and the number
of produced $B$ mesons, assuming equal production rates of charged and
neutral pairs. 
In Figure \ref{fig:Cchisq} the Likelihood function for the two modes is plotted.

Table \ref{t:results} shows for both decay chains the branching fraction
we measure, together with the quantities entering into its computation.
The statistical error on the number of events is taken as the shift from
the central value that changes the quantity $\chi^2\equiv -2\ln{\cal L}$
by one unit. We also give the statistical significance $S$, computed as the
square root of the difference between the value of $\chi^2$ for zero
signal and the value at its minimum. 

\providecommand{\corrEffB}{Corr. $\epsilon\times\prod\calB_i$ (\%)}
\providecommand{\signifSyst}{Signif. w syst. ($\sigma$)}
\providecommand{\bfemsix}{${\cal B}(\times10^{-6}$)}
\begin{table*}[htbp]
\caption{
signal event yield with statistical uncertainty, detection efficiency
($\epsilon$, \%), daughter branching fractions (\%),
significance $S$, and branching fraction result for each decay chain.
}
\label{t:results}
\begin{center}
\begin{tabular}{lcccccc}
\dbline
Mode                    & Signal yield          & $\epsilon$    & $\prod\calB_i$        & $S$           & \bfemsix\ (CL 90 \% )          \\
\sgline
\quad\fetaKstz        & $21 \pm 6$  & 19.0            & 26.1                 & 5.4            & $19.8^{+6.5}_{-5.6} \pm 1.7$  \\
\quad\fetaKstp        &$14 \pm 7$ & 17.6            & 17.9                  & 3.2            & $22.1^{+11.1}_{-9.2} \pm 3.3$ (33.9)\\
\dbline
\end{tabular}
\end{center}
\end{table*}

\begin{figure}[htbp]
\begin{center}
 \psfiletwoBB{0 0 425 425}{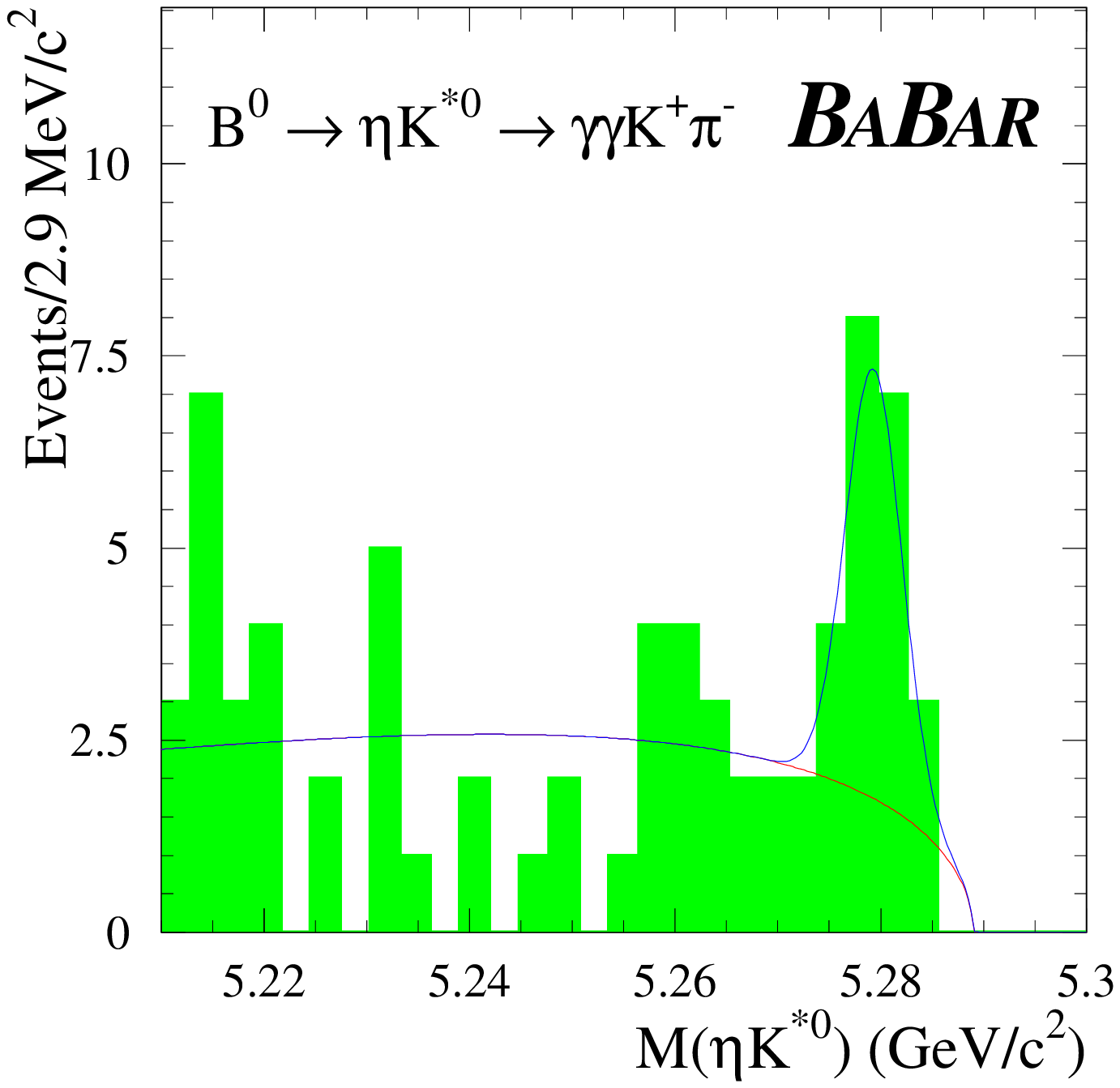}{0 0 425 425}{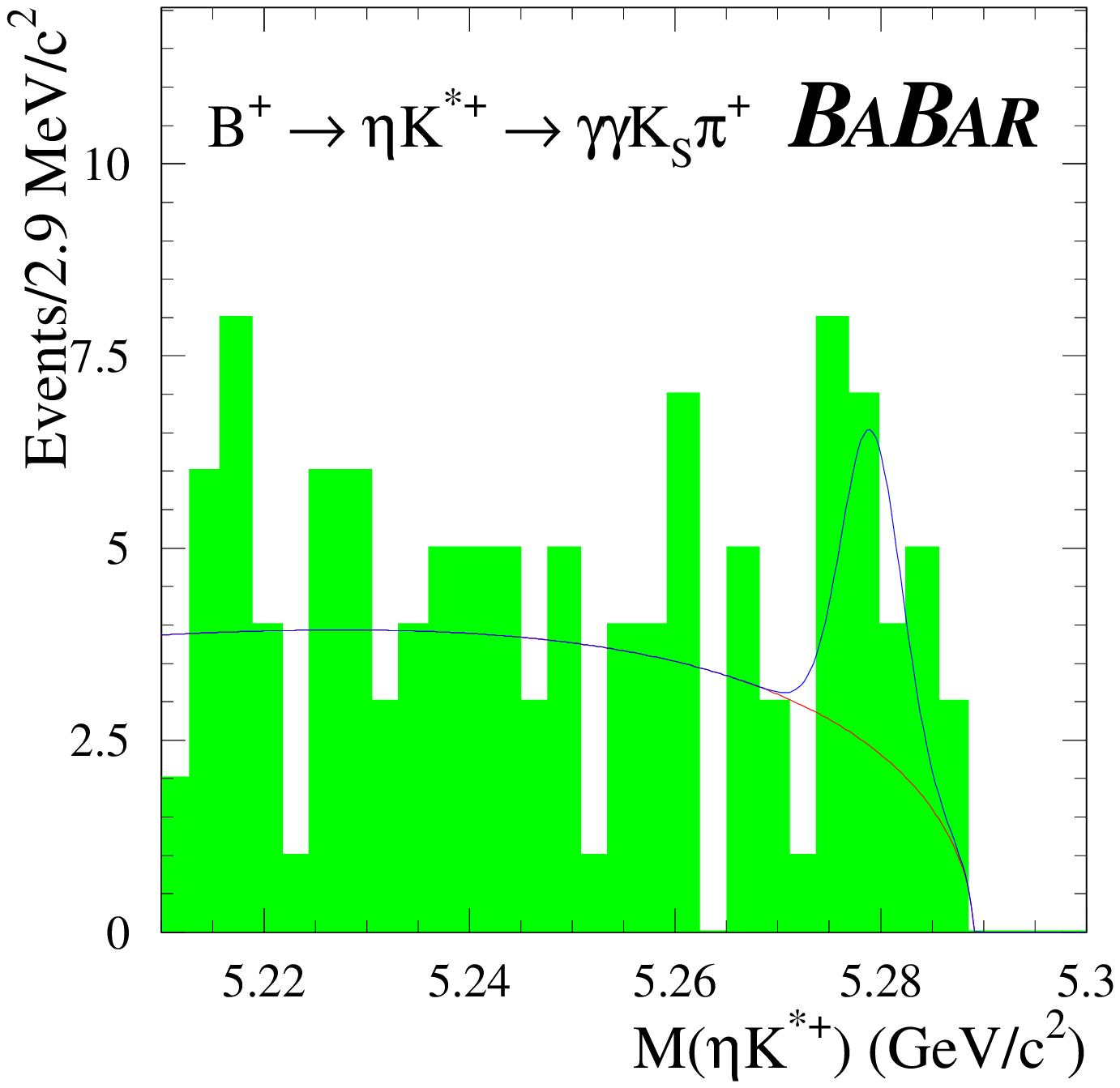}{0.99}
\end{center}
\vspace{1.5cm}
 \caption{\label{fig:projMb}
$B$ candidate invariant mass for \etaKstz (left) and \etaKstp (right).
Histograms represent data, and smooth curves represent the fit function.  }
\end{figure}

In Fig.\ \ref{fig:projMb}\ we show projections of \mec for both modes.
The projections are made by applying a requirement on
the individual event likelihood (computed without \mec) to select the
more signal-like events.  The overlaid curves represent the ML fit PDF
scaled to take into account the effect of the additional requirement.

For each measurement the supporting requirement-based analysis yielded
compatible results with comparable, if somewhat larger, statistical
errors.

\section{Systematic studies}
\label{sec:Systematics}

We have evaluated systematic errors, which are dominated
by the PDF uncertainties (6--12\%, depending on the decay mode).
To determine these we varied parameters of the PDFs
within their uncertainties and estimated the impact on the fit yield.
This is the only additive systematic error; all others are
multiplicative. 

Auxiliary studies lead to systematic errors of 1\%, 2.5\%, and 5\%\
respectively for the imperfect simulation of track, photon, and \KS\
efficiencies.  These errors are summed linearly for the $B$ daughters.
The $B$ production systematic error has
been estimated in a separate study to be 1.6\%.  Published
world averages
\cite{pdg}\ provide the $B$ daughter branching fraction uncertainties.

Systematic errors associated with the event selection are minimal given
the generally loose requirements.  We account explicitly for $|\cos{\theta_T}|$
(1\%), for which we observe a nearly uniform distribution in the signal
simulation.  We also include errors of 4\%\ due to the PID requirements.

\section{Summary}
\label{sec:Summary}

We have found significant event yields in the decay \etaKst,  as reported in Table \ref{t:results}.
The final results are generally in agreement with those previously
reported \cite{CLEOetapr}.  We confirm the rather larger than predicted
\cite{thy}\ rate for \etaKst\ obtained by the CLEO Collaboration \cite{CLEOetapr}.  
The enhancement in \etaKst\ could be due to constructively
interfering internal penguin diagrams \cite{lipkin}.

\section{Acknowledgments}
\label{sec:Acknowledgments}

We are grateful for the 
extraordinary contributions of our \pep2\ colleagues in
achieving the excellent luminosity and machine conditions
that have made this work possible.
The collaborating institutions wish to thank 
SLAC for its support and the kind hospitality extended to them. 
This work is supported by the
US Department of Energy
and National Science Foundation, the
Natural Sciences and Engineering Research Council (Canada),
Institute of High Energy Physics (China), the
Commissariat \`a l'Energie Atomique and
Institut National de Physique Nucl\'eaire et de Physique des Particules
(France), the
Bundesministerium f\"ur Bildung und Forschung
(Germany), the
Istituto Nazionale di Fisica Nucleare (Italy),
the Research Council of Norway, the
Ministry of Science and Technology of the Russian Federation, and the
Particle Physics and Astronomy Research Council (United Kingdom). 
Individuals have received support from the Swiss 
National Science Foundation, the A. P. Sloan Foundation, 
the Research Corporation,
and the Alexander von Humboldt Foundation.


\begin{thebibliography}{99}

\bibitem{bib:conjugate} 
Charge conjugate states are implied throughout this paper.

\bibitem{CKM}
N.~Cabibbo, \jprl {\bf 10}, 531 (1963);
M.~Kobayashi and T.~Maskawa, \progtp {\bf 49}, 652 (1973).


\bibitem{babar}
The \babar\ Collaboration, B. \ Aubert {\em et al.},
SLAC-PUB-8569, hep-ex/0105044 (to appear in \nim).

\bibitem{pep} 
PEP-II Conceptual Design Report, SLAC-R-418 (1993).

\bibitem{KpiPRL}
The \babar\ Collaboration, B.\ Aubert {\em et al.},
SLAC-PUB-8838, hep-ex/0105061, see Fig. 1(b) (Submitted to \jprl).

\bibitem{geant}
The \babar\ detector Monte Carlo simulation is based on GEANT: \\
R.~Brun {\it et al.}, CERN DD/EE/84-1.

\bibitem{CLEO-fisher}
CLEO Collaboration,
D.M.~Asner {\it et al.},
\jprd\ {\bf 53}, 1039 (1996).

\bibitem{cbshape}
T.~Skwarnicki  [Crystal Ball Collaboration],
``A Study Of The Radiative Cascade Transitions Between The Upsilon-Prime And Upsilon Resonances,''
DESY F31-86-02 (thesis, unpublished) (1986).

\bibitem{argus}
{ARGUS Collaboration, H. Albrecht {\em et al.}, \plb\ {\bf 241} (1990) 278; \\
\plb\ {\bf 254} (1991) 288}.

\bibitem{pdg}
Particle Data Group, D.E.~Groom {\it et al.},
\epjc\ {\bf 15}, 1 (2000).

\bibitem{CLEOetapr}\label{ref:CLEOetapr}
CLEO Collaboration S.J.\ Richichi {\it et al.}, \jprl{85}, 520 (2000); \\
CLEO CONF 99-12 (1999).

\bibitem{thy}
A. Ali, G. Kramer, and C.D. L\"{u}, \jprd {\textbf 58}, 094009 (1998); \\
Y. H. Chen {\it et al.}, \jprd {\textbf 60}, 094014 (1999).

\bibitem{lipkin}\label{bib:lipkin}
H.\ J.\ Lipkin, \plb\ B \textbf{254}, 247 (1991).


\end{thebibliography}
\end{document}